\documentclass[%
reprint,
aps,
superscriptaddress,
nofootinbib,
nobibnotes,
]{revtex4-1}

\usepackage{xcolor}

\usepackage{graphicx}
\graphicspath{{./fig/}}

\usepackage{latexsym,amsfonts,amssymb,amsmath,makeidx,mathrsfs,bm,bbm}
\DeclareMathAlphabet{\mathpzc}{OT1}{pzc}{m}{it}

\usepackage{multirow,lineno}

\usepackage{verbatim,acronym}
\usepackage{subfigure}

\usepackage{epsfig}

\usepackage{slashed,skak}

\usepackage[english]{babel}

\usepackage{threeparttable,multirow,txfonts,makecell,tabularx}
\usepackage[%
colorlinks=true,
linkcolor=blue,
citecolor=blue,
]{hyperref}

\begin{document}

\title{Gravitational wave sources for Pulsar Timing Arrays}

\author{Ligong Bian}
\email{lgbycl@cqu.edu.cn}
\affiliation{Department of Physics and Chongqing Key Laboratory for Strongly Coupled Physics,
Chongqing University, Chongqing 401331, China}
\affiliation{Center for High Energy Physics, Peking University, Beijing 100871, China}

\author{Shuailiang Ge}
\email{sge@pku.edu.cn}
\affiliation{Center for High Energy Physics, Peking University, Beijing 100871, China}
\affiliation{School of Physics and State Key Laboratory of Nuclear Physics and Technology, Peking University, Beijing 100871, China}

\author{Jing Shu}
\email{jshu@pku.edu.cn}
\affiliation{School of Physics and State Key Laboratory of Nuclear Physics and Technology, Peking University, Beijing 100871, China}
\affiliation{Center for High Energy Physics, Peking University, Beijing 100871, China}
\affiliation{Beijing Laser Acceleration Innovation Center, Huairou, Beijing, 101400, China}

\author{Bo Wang}
\affiliation{International Centre for Theoretical Physics Asia-Pacific, University of Chinese Academy of Sciences, 100190 Beijing, China}

\author{Xing-Yu Yang}
\email{xingyuyang@kias.re.kr}
\affiliation{Quantum Universe Center (QUC), Korea Institute for Advanced Study, Seoul 02455, Republic of Korea}

\author{Junchao Zong}
\affiliation{Department of Physics, Nanjing University, Nanjing 210093, China}
\affiliation{CAS Key Laboratory of Theoretical Physics, Institute of Theoretical Physics, Chinese Academy of Sciences, Beijing 100190, China}


\begin{abstract}

Very recently, several pulsar timing array collaborations, including CPTA, EPTA, and NANOGrav, reported their results from searches for an isotropic stochastic gravitational wave background (SGWB), with each finding positive evidence for SGWB. In this work, we assessed the credibility of interpreting the Hellings-Downs correlated free-spectrum process of EPTA, PPTA, and NANOGrav as either the result of supermassive black hole binary mergers or various stochastic SGWB sources that originated in the early Universe, including first-order phase transitions, cosmic strings, domain walls, and large-amplitude curvature perturbations. Our observations show that the current new datasets do not display a strong preference for any specific SGWB source based on Bayesian analysis.

\end{abstract}

\maketitle

\section{ Introduction}

Pulsar timing array (PTA) experiments provide a unique window to probe the gravitational waves (GWs) at nano-Hertz frequencies, with possible sources being supermassive black hole binaries (SMBHBs)~\cite{Rajagopal:1994zj,Phinney:2001di,Jaffe:2002rt,Wyithe:2002ep,Arzoumanian:2020vkk}, curvature perturbations~\cite{Ananda:2006af,Baumann:2007zm}, 
and new-physics models including
first-order phase transition (FOPT)~\cite{Kosowsky:1992rz,Caprini:2010xv}, 
cosmic strings~\cite{Siemens:2006yp}, 
and domain walls~\cite{Hiramatsu:2013qaa}, etc.

Previously, hints of a stochastic common-spectrum process have been observed in the pulsar-timing datasets of NANOGrav~\cite{Arzoumanian:2020vkk}, PPTA~\cite{Goncharov:2021oub}, EPTA~\cite{Chen:2021rqp} and IPTA~\cite{Antoniadis:2022pcn}, which has aroused enormous interests in the communities of astrophysics, cosmology, and particle physics. Since then, 
numerous interpretations have been proposed for such a stochastic common-spectrum process in the literature based on different models, including SMBHBs~\cite{Vaskonen:2020lbd,DeLuca:2020agl,Kohri:2020qqd}, cosmic strings~\cite{Ellis:2020ena,Blasi:2020mfx,Samanta:2020cdk,Bian:2022tju}, FOPT~\cite{Nakai:2020oit,Addazi:2020zcj,Ratzinger:2020koh,Xue:2021gyq}, domain walls~\cite{Bian:2020urb,Ferreira:2022zzo}, and so on. 
Very recently, NANOGrav released their new 15-yr dataset~\cite{NANOGrav:2023gor,NANOGrav:2023hvm,NANOGrav:2023hde}, CPTA released their first data~\cite{Xu:2023wog},
EPTA release the second data~\cite{Antoniadis:2023ott}, and PPTA released their third data set~\cite{Reardon:2023gzh,Reardon:2023zen,Zic:2023gta}. Therein, NANOGrav, EPTA, and CPTA inspiringly show positive evidence for the detection of stochastic gravitational-wave background (SGWB) in the common-spectrum process. 

Compared to the previously observed common-spectrum process in the old NANOGrav 12.5-yr dataset and the results from other collaborations like EPTA and PPTA, this time we not only have robust evidence for the common-spectrum process, but also have positive evidence concerning the Hellings-Downs (HD) correlation, which provides direct evidence for the gravitational wave quadrupolar signal. In this Letter, we incorporate the new PTA datasets from the three collaborations - PPTA, EPTA, and NANOGrav. We employ the Bayesian analysis method to contrast interpretations between different SGWB models and fit each model separately. Ultimately, we find no strong evidence favoring any potential cosmological sources (such as SMBHBs, etc) from the early Universe.

\section{Models of generating SGWBs}

We discuss five main mechanisms in the early Universe that can generate SGWB at nano-Hertz frequencies, which are 1) SMBHB, 2) FOPT, 3) cosmic strings, 4) domain walls, and 5) large amplitude curvature perturbations.
We refer the readers to Ref.~\cite{Bian:2020urb} and references therein for a summary of these models.
For convenience, we have also summarized the GW spectra of the latter four models with corresponding references in the Appendix.

Centers of most galaxies likely host supermassive black holes, forming binary systems during galaxy mergers~\cite{Kormendy:2013dxa,EventHorizonTelescope:2019dse}.
These systems emit gravitational radiation, creating a Gravitational Wave Background (GWB) detectable in the PTA band.
The GWB's properties depend on the SMBHBs' characteristics and evolution.
For binaries purely evolving through GW emission, the power spectral density follows a power law with a spectral index of -13/3~\cite{Phinney:2001di}, influenced by interactions with the local galactic environment~\cite{Begelman:1980vb}.

FOPTs happening in the early Universe arise in many models beyond the Standard Model of particle physics. For example, they are usually associated with explaining the baryon asymmetry (see e.g., Ref.~\cite{Morrissey:2012db}). An FOPT can generate gravitational waves in multiple ways, including collisions of vacuum bubbles, relevant shocks in the plasma, sound waves in the plasma after bubble collisions, and the magnetohydrodynamic turbulence in the plasma after bubble collisions~\cite{Caprini:2015zlo}. Here, following Ref.~\cite{Bian:2020urb}, we consider the scenario that sound-wave contribution dominates. The GW spectrum is mainly determined by the latent heat $\alpha_{PT}$, the inverse time duration of FOPT $\beta$ which is usually rescaled by the Hubble parameter $H_n$ at the bubble nucleation temperature $T_n$ (which is approximately $T_*$, the temperature when the GW are produced), and the velocity of expanding bubble wall in the plasma background $v_b$~\cite{Caprini:2015zlo}.  

A cosmic string is a one-dimensional topological defect associated with a symmetry breaking, which is also predicted in many models beyond the Standard Model. Infinite strings will intersect and generate string loops~\cite{Vachaspati:1984dz}. The string loops can oscillate and vibrate to emit gravitational waves~\cite{Vachaspati:1984gt}. The strings can also develop the structure of kinks and cusps that can generate gravitational waves~\cite{Damour:2000wa, Damour:2001bk}. GWs emitted from cosmic strings mainly depend on the parameters $G\mu$ and $\alpha_{CS}$. $G$ is the Newton gravitational constant and $\mu$ is the string tension (energy per unit length). $\alpha_{CS}$ is the loop-size parameter representing the ratio of the loop size to the Hubble length (or more naturally, the correlation length~\cite{Sousa:2013aaa,Gouttenoire:2019kij}). Note we are discussing gauge strings associated with a gauge symmetry breaking, the energy of which is mainly lost in GWs, while the global strings associated with a global symmetry breaking lose energy mainly in the form of Goldstone bosons~\cite{Vilenkin:1986ku}.

A domain wall is another kind of topological defects, which is two-dimensional. It is formed when discrete degenerate vacua are present after a symmetry breaking, which also naturally arises in many beyond-Standard-Model theories. The evolution of the domain wall network can generate gravitational waves; see e.g., Ref.~\cite{Hiramatsu:2013qaa}. To avoid the domain wall problem that domain walls dominate the Universe, a bias potential $\Delta V$ is usually introduced to kill the domain wall network by explicitly breaking the vacua degeneracy~\cite{sikivie1982axions, gelmini1989cosmology, larsson1997evading}. $\Delta V$ determines the time when the network disappears and thus marks the location of GW spectrum's peak frequency. Another key factor is the domain wall tension $\sigma$, i.e., the energy per unit wall area. 

GWs can also be generated by the curvature perturbations due to the coupling at nonlinear order between scalar and tensor modes.
Via the coupling with the tensor modes, scalar perturbations can induce GWs; see e.g., Refs.~\cite{Ananda:2006af, Baumann:2007zm, Kohri:2018awv}.
The power spectrum of curvature perturbations is assumed to be a power law, $P_{\mathcal{R}}(k) \propto P_{\mathcal{R}0}(k/k_*)^{m}$, where $k$ is the wavenumber and $k_*$ is the wavenumber at the frequency around $1\mathrm{yr}^{-1}$. The corresponding GW spectrum is then $\Omega_{\mathrm{GW}}(k)\propto P_{\mathcal{R}}^2(k)$. The amplitude $P_{\mathcal{R}0}$ and the slope $m$ are the two key parameters that determine the GW spectrum.

\section{Comparisons among SGWB models}

By using the fitting results of the free spectrum with the HD correlation from the datasets of NANOGrav, PPTA and EPTA, we can make comparisons between the following models:
SMBHBs, FOPT, cosmic strings, domain walls, and scalar-induced GWs, which are labeled as $M_{i}$ ($i = 1, 2, 3, 4, 5$) in sequence, respectively. We list the bayesian prior range of the model parameters in table. \ref{PPT} and 
the results are summarized in Eqs.~(\ref{eqB})-(\ref{eqB_epta}).
In addition, the corresponding interpretation of Bayes factors is shown in Table~\ref{tab:Bayes_factor_intp}. 

EPTA and NANOGrav have a weak evidence while PPTA has a positive evidence in favor of the cosmic strings and FOPT explanations against SMBHBs.
A positive evidence in favor of the SMBHB, cosmic strings, scalar-induced GWs, FOPT against domain walls is shown in all the datasets of EPTA, PPTA and NANOGrav.
Especially, NANOGrav and PPTA show more inclination in favor of the FOPT explanation than the other sources, while EPTA are more sensitive to cosmic-string explanation.

Upon comparing these explanations, we find that none of the above models has a distinct advantage over others in interpreting the common-spectrum process with the HD correlation implied in the datasets of NANOGrav, PPTA and EPTA.

\begin{table}[htpb]
    \caption[]{%
        Bayes factors can be interpreted as follows:
        for comparing a candidate model $M_i$ against another model $M_j$, a Bayes factor of 20 corresponds to a belief of 95\% in the statement ``$M_{i}$ is true'', which means a strong evidence in favor of $M_{i}$~\cite{10.2307/2291091}.
    }
    \label{tab:Bayes_factor_intp}
    \begin{tabular}{lr}
        \hline
        $B_{ij}$ &  Evidence in favor of $M_{i}$ against $M_{j}$ \\
        \hline
        $1-3$ & Weak \\
        $3-20$ & Positive\\
        $20-150$ & Strong\\
        $\ge 150$ & Very strong \\
        \hline
    \end{tabular}
\end{table}

\begin{equation}\label{eqB}
   B_{ij}^{\textrm{NG15}}= 
\begin{pmatrix}
    1 & 0.49 &0.55&5.19&1.34\\
    2.03 & 1 & 1.12&10.55 & 2.72\\
    1.82 & 0.90 & 1 & 9.46 & 2.44\\
    0.19&  0.09&0.11 &1 & 0.26\\
    0.75 & 0.37 & 0.41 & 3.88& 1
\end{pmatrix}
\end{equation}

\begin{equation}\label{eqB_ppta}
   B_{ij}^{\textrm{PPTA}}= 
\begin{pmatrix}
    1 & 0.27 &0.32& 2.53 & 0.58\\
    3.64 & 1 & 1.16& 9.2 & 2.10\\
    3.13 & 0.86 & 1 & 7.92 & 1.81\\
    0.40&  0.11&0.13 &1 & 0.23\\
    1.73 & 0.48 & 0.55 & 4.37& 1
\end{pmatrix}
\end{equation}

\begin{equation}\label{eqB_epta}
   B_{ij}^{\textrm{EPTA}}= 
\begin{pmatrix}
    1 & 0.67 &0.47 &6.87 &1.65\\
    1.50 & 1 & 0.70&10.30 & 2.47\\
    2.15 & 1.43 & 1 & 14.75 & 3.53\\
    0.15&  0.10&0.07 &1 & 0.24\\
    0.61 & 0.41 & 0.28 & 4.18& 1
\end{pmatrix}
\end{equation}

\section{Constraints on SGWB models}

\begin{figure}[htbp]
    \centering
    \includegraphics[width=0.4\textwidth]{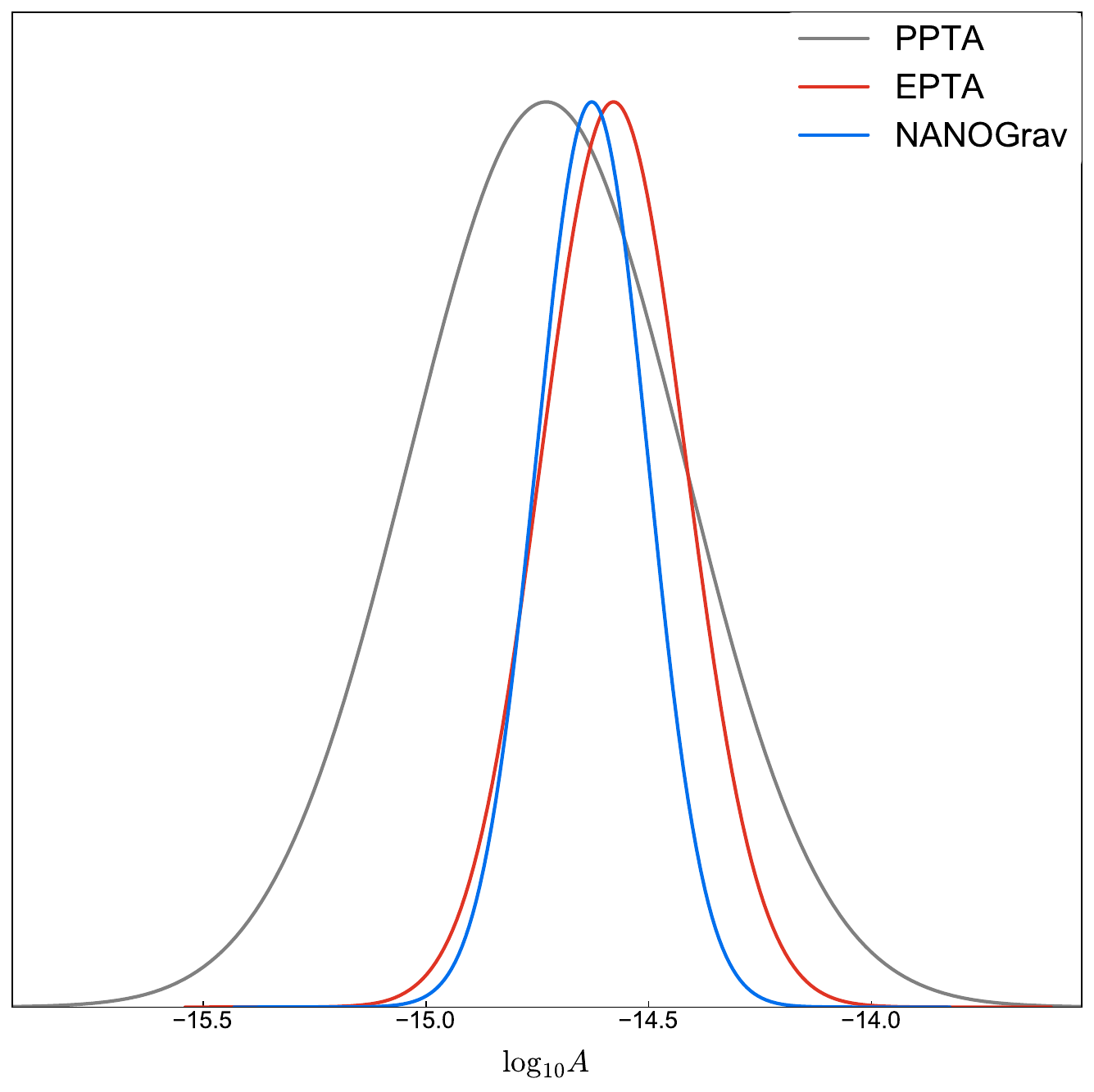}
    \caption{Constraints on the SMBHB parameter $\log_{10}A$ from the Bayesian model fitting.}
    \label{SMBHB}
\end{figure}

In Fig.~\ref{SMBHB}, we show the constraints on the log-amplitude $\log_{10}A$ of the SMBHB power-law spectrum. Analyses of PPTA, EPTA, and NANOGrav datasets yield the results $\log_{10}A \sim [-15.04,-14.42]$, $[-14.74,-14.42]$, and $[-14.76,-14.50]$ at 68\% confidence level (C.L.), respectively.

\begin{figure}[htbp]
    \centering
    \includegraphics[width=0.45\textwidth]{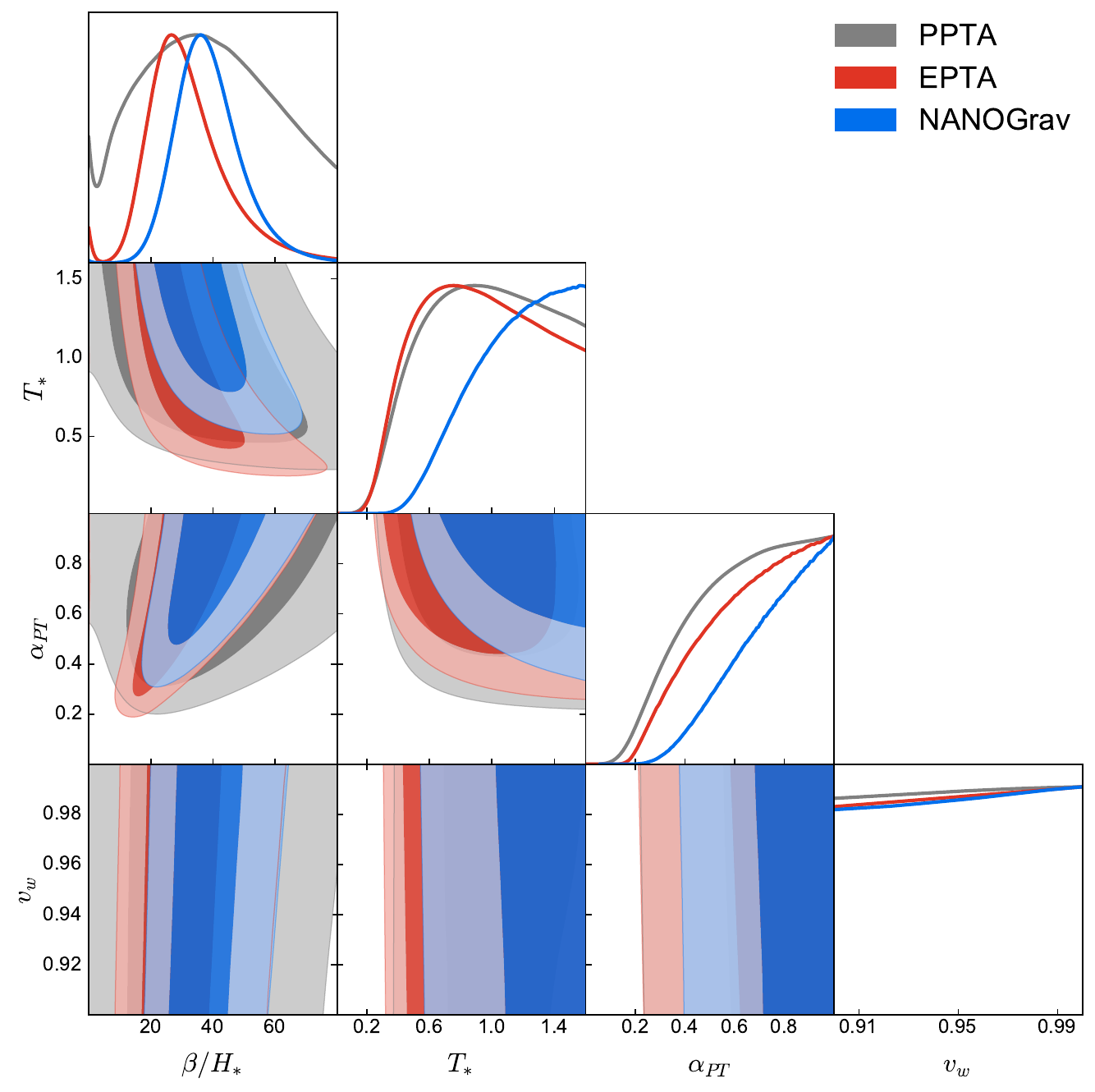}
    \caption{The constraints on parameters of FOPT from Bayesian model fitting. Contours contain 68\% and 95\% of the probability.}
    \label{PTGW}
\end{figure}

Fig.~\ref{PTGW} shows the result for the FOPT case from the Bayesian model fitting. 
The data constraint based on the PPTA dataset favors a moderate latent heat $\alpha_{PT}\geq 0.548$ and a duration $\beta/H_{*}\sim [9,59]$ at the phase transition temperature $T_* \sim [0.61, 1.33]$~MeV at $68\%$ C.L. Likewise, EPTA dataset at the same confidence level favors a latent heat $\alpha_{PT}\geq 0.591$, accompanied by a duration $\beta/H_{*}\sim [22,40]$ at $T_* \sim [0.48, 1.30]$~MeV. Finally, the NANOGrav dataset at the same confidence level favors $\alpha_{PT}\geq 0.692$, $\beta/H_{n}\sim [29,47]$, and $T_n \geq 1.03$~MeV. 
Noting that the energy injection from the phase transition would change the BBN and CMB observations~\cite{Bai:2021ibt, Deng:2023seh}, which excludes some slow and strong phase transitions around $T_* \sim 1$~MeV.

\begin{figure}[htbp]
    \centering
    \includegraphics[width=0.45\textwidth]{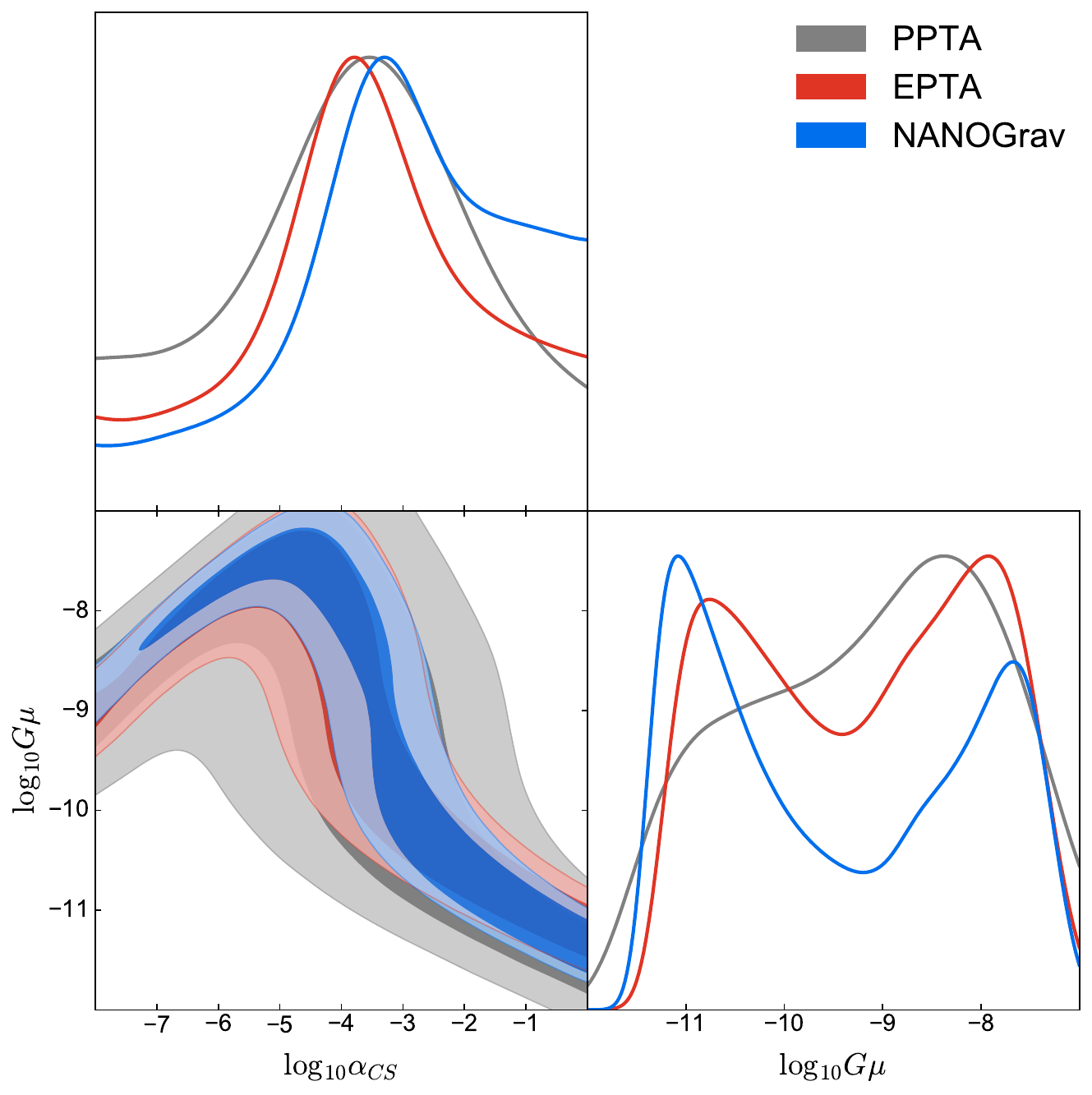}
    \caption{ The constraints on parameters of cosmic strings from the Bayesian model fitting. Contours contain 68\% and 95\% of the probability.}
    \label{CSGW}
\end{figure}

The results based on the Bayesian model fitting for the case of cosmic string network are shown in Fig.~\ref{CSGW}. The constraints yield $\log_{10} G\mu\sim[-10.2,-7.5],\ [-10.4,-8.0]$, and $[-10.9, -8.1]$ at 68\% C.L. under PPTA, EPTA, and NANOGrav datasets, respectively, implying a $U(1)$ symmetry-breaking scale $\eta\sim \mathcal{O}(10^{13-14})$ GeV of local strings.
Meanwhile, we also obtain constraints on the loop-size parameter $\alpha_{CS}$ that $\log_{10} \alpha_{CS}\sim [-5.5,-1.5],\ [-5.3,-1.5]$, and $[-4.4,-0.7]$ at 68\% C.L. from PPTA, EPTA, and NANOGrav datasets, respectively, which are well below the typical value of $\alpha_{CS}=0.1$ suggested by simulations~\cite{Blanco-Pillado:2013qja,Blanco-Pillado:2017oxo}.

\begin{figure}[t]
    \centering
    \includegraphics[width=0.45\textwidth]{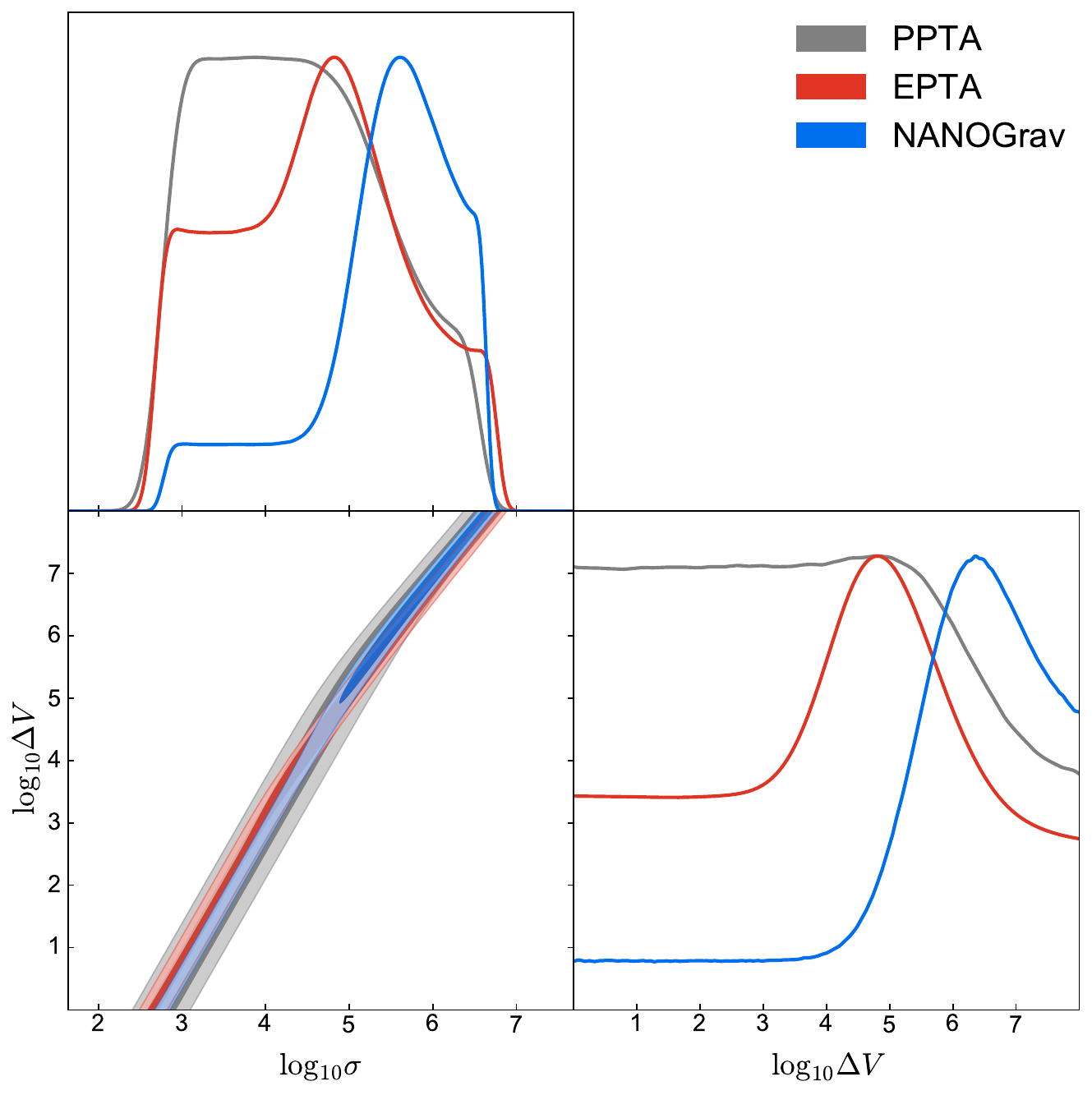}
    \caption{The constraints on parameters of domain wall from Bayesian model fitting. Contours contain 68\% and 95\% of the probability.}
    \label{DWGW}
\end{figure}

In Fig.~\ref{DWGW}, we show the results for the domain-wall case based on the Bayesian model fitting. At $68\%$ C.L., we get tight bounds on the bias $\Delta V$ and the surface energy density $\sigma$:
$\log_{10} (\sigma/\mathrm{TeV}^{3})\sim[2.98,5.12]$, $\log_{10} (\Delta V/\mathrm{MeV}^{4}) \leq 4.94$ under the PPTA dataset,
$\log_{10} (\sigma/\mathrm{TeV}^{3})\sim[2.89,5.56]$, $\log_{10} (\Delta V/\mathrm{MeV}^{4})\leq5.26$ under the EPTA dataset, and $\log_{10} (\sigma/\mathrm{TeV}^{3})\sim[5.11,6.50]$, $\log_{10} (\Delta V/\mathrm{MeV}^{4})\sim[5.50,7.95]$ under the NANOGrav dataset. 
Thus, considering a $Z_2$ domain wall network as an example, the results imply that the symmetry breaking scale should be $\eta \lesssim 10^{4}$~TeV for $\sigma= 2\sqrt{2\lambda} \eta^3/3$ assuming the interaction coupling as $\lambda\sim  \mathcal{O}(10^{-2})$.

\begin{figure}[t]
	\centering
	\includegraphics[width=0.45\textwidth]{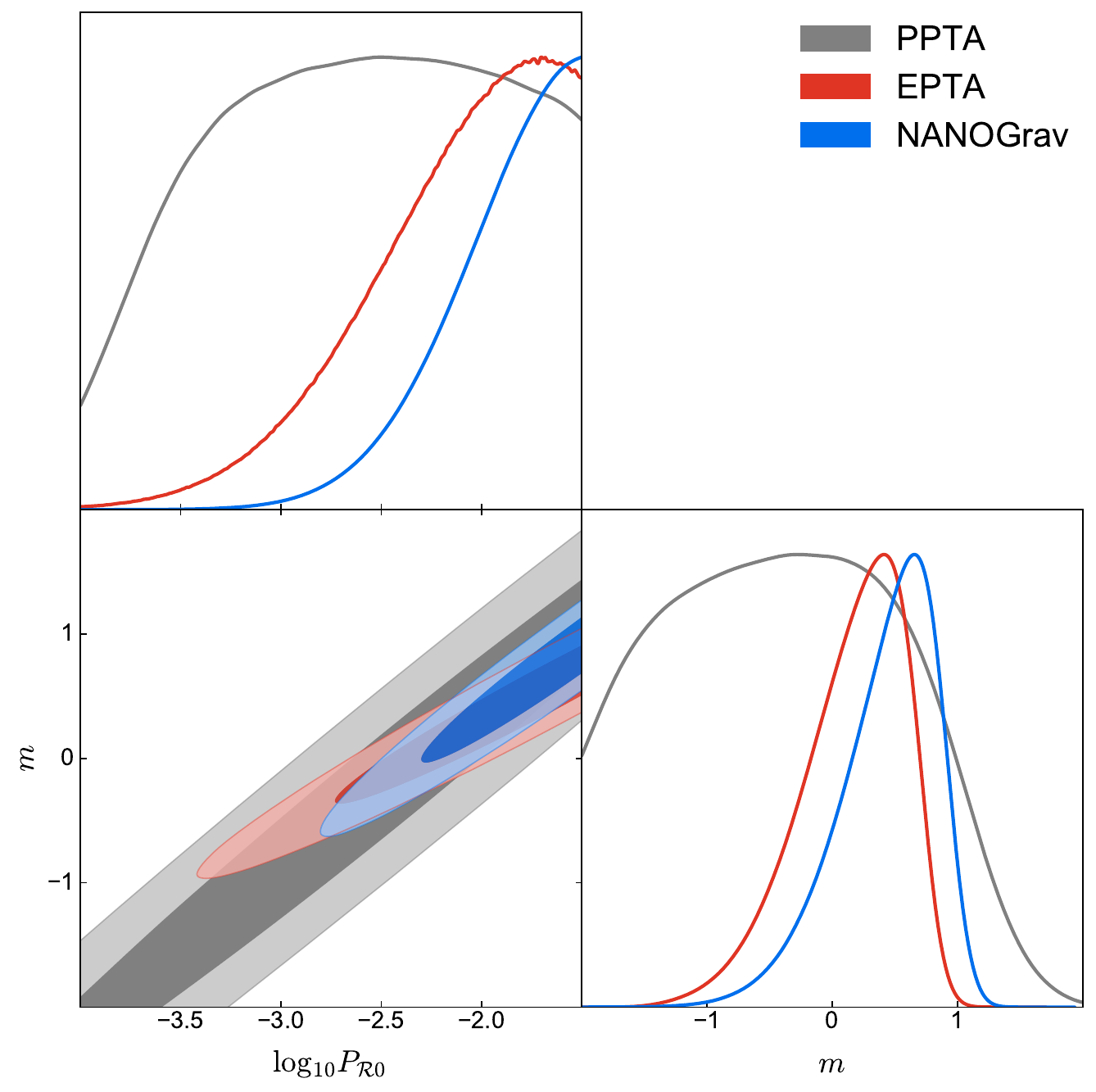}
	\caption{The constraints on parameters of power spectrum of curvature perturbations from the Bayesian model fitting. Contours contain 68\% and 95\% of the probability.}
    \label{IGW}
\end{figure}

In the case of scalar-induced GWs, we find
$\log_{10}P_{\mathcal{R} 0} \sim [-3.17,-1.67]$ and $m \sim [-1.27,0.53]$ are allowed by the PPTA dataset,
$\log_{10}P_{\mathcal{R} 0} \geq -2.32$ and $m \sim [-0.12,0.68]$ are allowed by the EPTA dataset,
$\log_{10}P_{\mathcal{R} 0} \geq -2.03$ and $m \sim [0.19,0.91]$ are allowed by the NANOGrav dataset at $68\%$ C.L.,
as shown in Fig.~\ref{IGW}.
The slope $m$ has a negative best-fit value from the PPTA dataset, which is consistent with the result from the old 12.5-yr NANOGrav dataset.
However, positive best-fit values of $m$ are obtained from the new datasets of EPTA and NANOGrav.
None of these results shows a $k^{3}$ slope which was suggested as a universal infrared behavior of GW spectrum~\cite{Cai:2019cdl}.
The large amplitude curvature perturbations are also related to the formation of primordial black holes (PBHs), which are attractive dark matter candidates and are also the possible sources for the merger events of black hole binaries~\cite{Sasaki:2016jop,Sasaki:2018dmp,Carr:2020gox}.
The best-fit value of amplitude $P_{\mathcal{R}0}$ from PPTA is similar to that from the old 12.5-yr NANOGrav dataset. In comparison, the new datasets of NANOGrav and EPTA give a larger best-fit value.
The larger best-fit amplitude from new datasets implies a larger corresponding PBH abundance which can be even larger if the non-Gaussianity of curvature perturbations is considered~\cite{Cai:2018dig,Bartolo:2018rku,Cai:2019elf}.

\section{ Conclusion and discussion}

We consider different SGWB sources as the possible interpretations of the strong stochastic common-spectrum process with HD correlation observed by the NANOGrav, PPTA and EPTA collaborations.
A Bayesian model comparison is carried out by fitting with the first 5 low-frequency bins of their HD free-spectrum data. Our results show that 
the current datasets from the three collaborations are all unable to distinguish one SGWB model as being obviously superior to the others.
We also place constraints on the parameter spaces of SMBHBs, FOPT, cosmic strings, domain walls, and curvature fluctuations, some of which can be further used to constrain the related new physics. 
Our study mainly indicates that:
1) parameter spaces of the slow phase transition are with moderate strength around 1~MeV scale, which can be further constrained by BBN and CMB observations;
2) cosmic strings are formed after the $U(1)$ symmetry breaking scale around $\eta\sim \mathcal{O}(10^{13-14})$ GeV;
3) the discrete symmetry breaking scale should be lower than $10^4$ TeV;
4) the PBHs from curvature perturbations are severely constrained.
In addition, we find that compared to EPTA and PPTA, the current data from NANOGrav can place much stronger constraints on SGWB model parameters.

Since all the cosmological SGWB models can reproduce the HD signal observed in the current datasets, more data from pulsar timing array are necessary to distinguish these models from SMBHBs. We further note that, more accurate GW spectrum based on numerical simulations and more accurate theoretical prediction of GW model parameters based on particle physics (such as the symmetry breaking scale for phase transitions, cosmic string, and domain wall) are also definitely crucial to settle down conclusively the preferred SGWB model(s).

\begin{acknowledgments}
This work is supported by the National Key Research and Development Program of China under Grant No. 2020YFC2201501 and 2021YFC2203004.
L.B. is supported by the National Natural Science Foundation of China (NSFC) under Grants No. 12075041 and No. 12147102, the Fundamental Research Funds for the Central Universities of China under Grants No. 2021CDJQY-011 and No. 2020CDJQY-Z003.
S.G. is supported by NSFC under Grant No. 12247147, the International Postdoctoral Exchange Fellowship Program, and the Boya Postdoctoral Fellowship of Peking University.
J.S. is supported by Peking University under startup Grant No. 7101302974 and the National Natural Science Foundation of China under Grants No. 12025507, No.12150015; and is supported by the Key Research Program of Frontier Science of the Chinese Academy of Sciences (CAS) under Grants No. ZDBS-LY-7003 and CAS project for Young Scientists in Basic Research YSBR-006.
XYY is supported in part by the KIAS Individual Grant QP090701.
\end{acknowledgments}

\bibliography{citelib}

\clearpage
\appendix

\section{GW spectrum from different models}

\subsubsection{First-order phase transition (FOPT)}

\noindent \textit{GW spectrum}~\cite{Caprini:2015zlo}:
\begin{eqnarray}
&&\Omega_{\mathrm{GW}}^{\rm sw}(f) h^2  \nonumber \\
&&=2.65 \times 10^{-6}(H_*\tau_{\rm sw})\left(\frac{\beta}{H_*}\right)^{-1} v_b
\left(\frac{\kappa_\nu \alpha_{PT} }{1+\alpha_{PT} }\right)^2\nonumber
\\
&&\times \left(\frac{g_*}{100}\right)^{-\frac{1}{3}}
\left(\frac{f}{f^{\rm sw}_{\rm peak}}\right)^3 \left[\frac{7}{4+3 \left(f/f^{\rm sw}_{\rm peak}\right)^2}\right]^{7/2}\,
\end{eqnarray}
with the peak frequency~\cite{Hindmarsh:2013xza,Hindmarsh:2015qta,Hindmarsh:2017gnf}: 
\begin{equation}
f^{\rm sw}_{\rm peak}=1.9 \times 10^{-5} \frac{\beta}{H_*} \frac{1}{v_b} \frac{T_*}{100}\left({\frac{g_*}{100}}\right)^{\frac{1}{6}} {\rm Hz }.
\end{equation}

\noindent \textit{Prior parameters}~\cite{Caprini:2015zlo}:
\begin{itemize}
    \item $\alpha_{PT}$, the latent heat;
    \item $\beta/H_n$, the inverse time duration of FOPT rescaled by $H_n$ which is the Hubble parameter at the nucleation temperature, $T_n$; 
    \item $T_*$, the temperature when GWs are produced. Approximately, we have $T_{*}\approx T_n$; 
    \item $v_b$, the velocity of vacuum bubble wall in the plasma background;
\end{itemize}

\noindent \textit{Other parameters}~\cite{Caprini:2015zlo, Ellis:2019oqb,Ellis:2020awk,Caprini:2019egz,Ellis:2018mja,Guo:2020grp}:
\begin{itemize}
    \item $\tau_{\rm sw}={\rm min} \left[\frac{1}{H_*},\frac{R_*}{\bar{U}_f}\right]$, the duration of the sound-wave period. $H_*$ is the Hubble parameter at $T_*$;
    \item $R_{*}$, the mean bubble separation. The following equality holds, $H_*R_*=v_b(8\pi)^{1/3}(\beta/H_*)^{-1}$;
    \item $\bar{U}_f^2\approx\frac{3}{4}\frac{\kappa_\nu\alpha}{1+\alpha}$, the root-mean-square (RMS) fluid velocity. $\alpha = \alpha_{PT}(1-\kappa_{\rm col})$ where $\kappa_{\rm col}\equiv E_{\rm wall}/E_{V}$ is the proportion of the vacuum energy to accelerate the bubble wall~\cite{Ellis:2019oqb}; 
    \item $\kappa_\nu$, the proportion of the vacuum energy transferred into the bulk fluid motion. It can be obtained via the hydrodynamic treatment of the plasma fluid~\cite{Espinosa:2010hh}.
    \item $g_*$, the effective number of degrees of freedom of relativistic species in the cosmic plasma at the time of GW formation.
\end{itemize}

\subsubsection{Cosmic strings}

\noindent \textit{GW spectrum}~\cite{Cui:2018rwi, Gouttenoire:2019rtn,Gouttenoire:2019kij}:

\begin{equation}
    \Omega_{\rm GW}^{\rm cs}(f) =\sum_k \Omega_{\rm GW}^{(k)}(f)\; 
\end{equation}
where for each $k$ mode,
\begin{eqnarray}\label{eq:GWdensity2}
\Omega_{\rm GW}^{(k)}(f) =&&
\frac{1}{\rho_c}
\frac{2k}{f}
\frac{\mathcal{F}_{\alpha}\,\Gamma^{(k)}G\mu^2}
{\alpha_{CS}\left( \alpha_{CS}+\Gamma G\mu\right)}
\int_{t_F}^{t_0}\!d\tilde{t}\;
\frac{C_{\rm eff}(t_i^{(k)})}{t_i^{(k)\,4}}\nonumber\\
&&\times\bigg[\frac{a(\tilde{t})}{a(t_0)}\bigg]^5\bigg[\frac{a(t^{(k)}_i)}{a(\tilde{t})}\bigg]^3\Theta(t_i^{(k)} - t_F)~.~~~~
\end{eqnarray}

\noindent \textit{Prior parameters}:
\begin{itemize}
    \item $G\mu$, where $G$ is the gravitational constant and $\mu$ is the string tension;
    \item $\alpha_{CS}$, the loop size parameter.
\end{itemize}

\noindent \textit{Other parameters}:
\begin{itemize}
     \item $\rho_c = 3H_0^2/8\pi G$, the critical density of the Universe;
     \item $\mathcal{F}_{\alpha}$ characterizes the proportion of the energy released by long strings, which we take as $0.1$;
     \item $C_{\rm eff}$ characterizes the loop production efficiency, which we take as $5.4$ and $0.39$ for the radiation- and matter-dominated Universe respectively~\cite{Gouttenoire:2019kij};
     \item $\Gamma\approx50$ is the gravitational emission efficiency of loops~\cite{Blanco-Pillado:2017oxo};
     \item $
\Gamma^{(k)} = \frac{\Gamma k^{-\frac{4}{3}}}{\sum_{m=1}^{\infty} m^{-\frac{4}{3}} } \;
$ are the Fourier modes of cusps~\cite{Olmez:2010bi, Blanco-Pillado:2013qja,Blanco-Pillado:2017oxo}; where $\sum_{m=1}^{\infty} m^{-\frac{4}{3}} \simeq 3.60$ and
$\sum_k \Gamma^{(k)}=\Gamma$.
\item  The formation time of loops of the $k$-mode: \\
\begin{equation}\label{eq:ti}
t_i^{(k)}(\tilde{t},f) = \frac{1}{\alpha_{CS}+\Gamma G\mu}\left[
\frac{2 k}{f}\frac{a(\tilde{t})}{a(t_0)} + \Gamma G\mu\;\tilde{t}\;
\right]\,.
\end{equation}
where $\tilde{t}$ is the GW emission time;
\item $t_F$, the formation temperature, after which cosmic string network reaches an attractor scaling regime;
\item If cusps dominate the small-scale structure of loops, the high mode and the low mode are related as:
$\Omega_{\rm GW}^{(k)}(f)
= \frac{\Gamma^{(k)}}{\Gamma^{(1)}}\,\Omega_{\rm GW}^{(1)}(f/k)
=k^{-4/3}\,\Omega_{\rm GW}^{(1)}(f/k)$\;.
 \end{itemize}

\subsubsection{Domain walls}

\noindent \textit{GW spectrum}~\cite{Hiramatsu:2013qaa,Kadota:2015dza,Zhou:2020ojf}:

\begin{equation}
\Omega^{\rm dw}_{\rm GW}(f) h^{2}=\Omega^{\rm peak}_{\rm GW} h^{2} S^{\rm dw}(f)
\end{equation}
where the peak GW amplitude is
\begin{eqnarray}
\Omega^{\mathrm{peak}}_{\mathrm{GW}} h^{2} &\simeq& 5.20 \times 10^{-20} \times \tilde{\epsilon}_{\mathrm{gw}} \mathcal{A}^{4}\left(\frac{10.75}{g_{*}}\right)^{1 / 3}\nonumber\\
&&
\times\left(\frac{\sigma}{1 \mathrm{TeV}^{3}}\right)^{4}
\left(\frac{1 \mathrm{MeV}^{4}}{\Delta V}\right)^{2}\,,\label{eq:gwdw}
\end{eqnarray}
and the shape function is 
\begin{eqnarray}
     S^{\rm dw}(f) &=& (f/f^{\rm dw}_{\rm peak})^3,
    ~~ f<f^{\rm dw}_{\rm peak} \\
     S^{\rm dw}(f) &=& (f/f^{\rm dw}_{\rm peak})^{-1},
    ~~ f\geq f^{\rm dw}_{\rm peak}
\end{eqnarray}
with the peak frequency~\cite{Hiramatsu:2013qaa}
\begin{equation}
f^{\rm dw}_{\mathrm{peak}}\simeq 3.99 \times 10^{-9} \mathrm{Hz}~ \mathcal{A}^{-\frac{1}{2}}\left(\frac{1 \mathrm{TeV}^{3}}{\sigma}\right)^{\frac{1}{2}}\left(\frac{\Delta V}{1 \mathrm{MeV}^{4}}\right)^{\frac{1}{2}}\;.
\end{equation}

\noindent \textit{Prior parameters}:
\begin{itemize}
    \item $\sigma$, the domain wall tension;
    \item $\Delta V$, the bias potential that explicitly breaks the degeneracy of vacua which makes the domain wall to disappear and thus marks the location of peak frequency.
\end{itemize}

\noindent \textit{Other parameters}:
\begin{itemize}
    \item $\mathcal{A}=1.2$, the area parameter whose value can be obtained via numerical simulations~\cite{Hiramatsu:2013qaa, Kadota:2015dza};
    \item $\tilde{\epsilon}_{\mathrm{gw}}=0.7$, the efficiency parameter of generating GWs~\cite{Hiramatsu:2013qaa}.
\end{itemize}

\begin{figure}[t]
    \centering
    \includegraphics[width=0.45\textwidth]{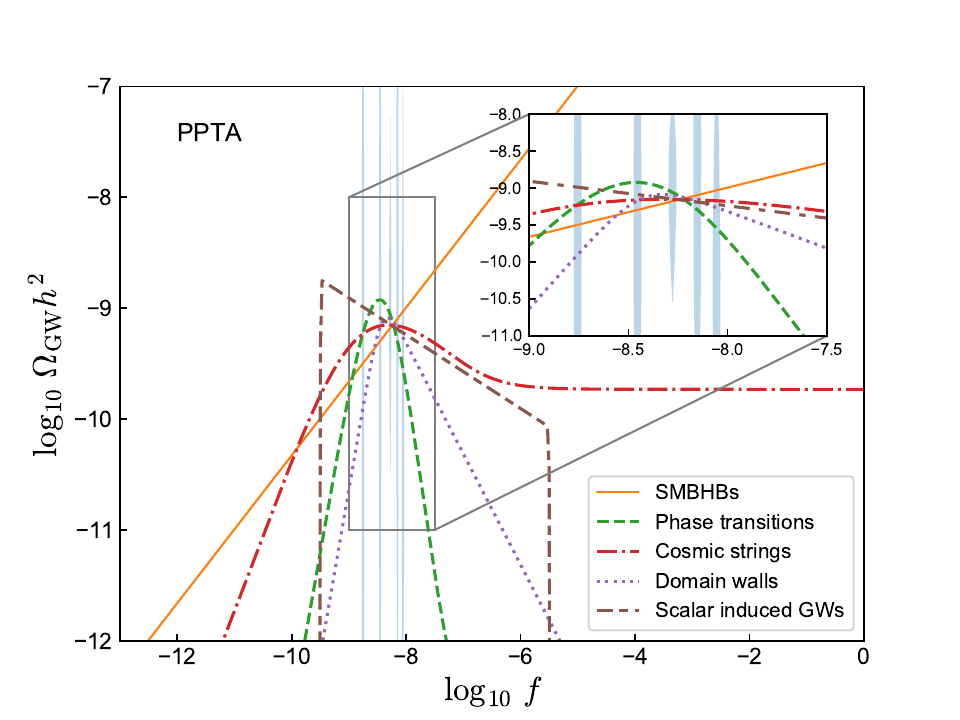}\\
    \includegraphics[width=0.45\textwidth]{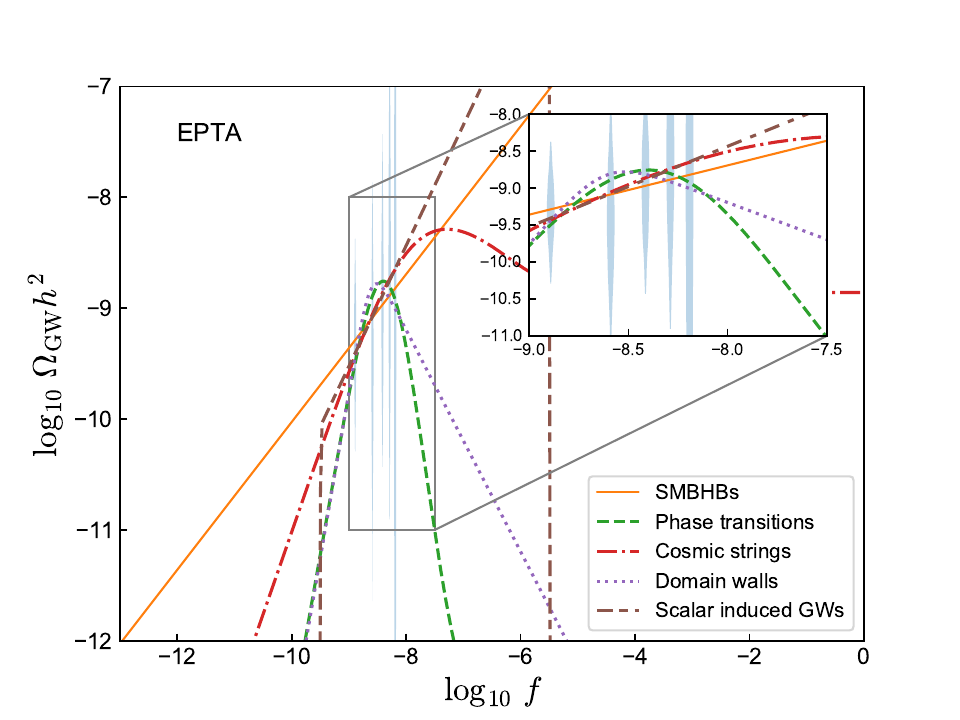}\\
     \includegraphics[width=0.45\textwidth]{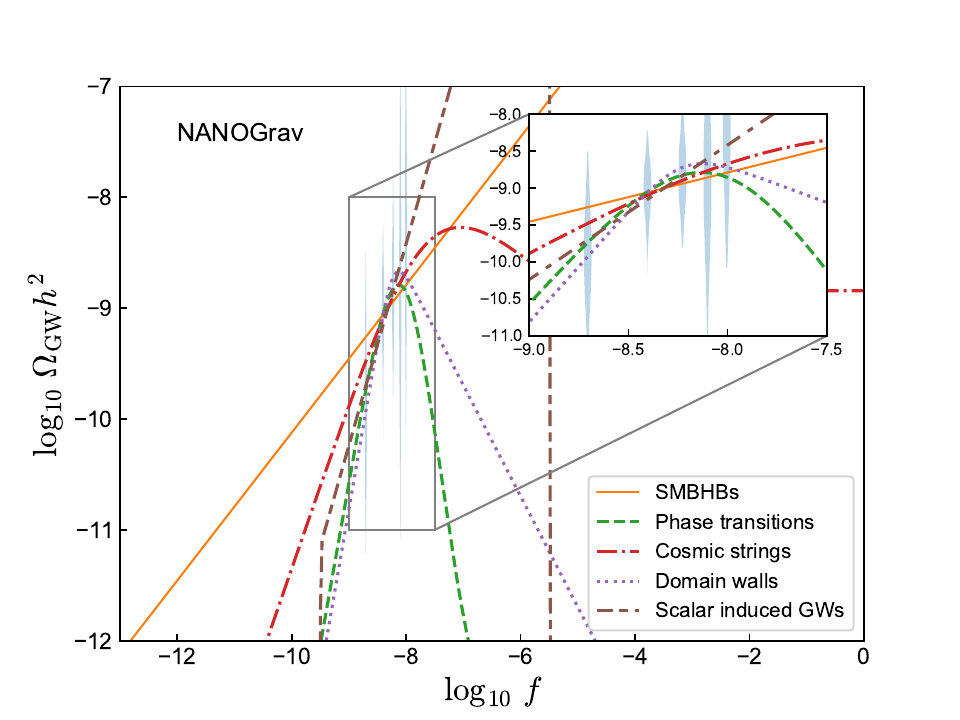}
    \caption[]{The GW energy spectra derived from different models employing the best-fit parameters from current new datasets of PPTA, EPTA and NANOGrav.
    The violin plots represent the first five frequency bins of the corresponding datasets.
    }
    \label{fig:median_fit}
\end{figure}

\subsubsection{Scalar-induced GWs}

\noindent \textit{GW spectrum}~\cite{Kohri:2018awv}:
\begin{equation}
	\begin{split}
		\Omega_{\mathrm{GW}}^{\rm si}(f)h^{2}
		&= \dfrac{1}{12}\Omega_{\mathrm{rad}}h^{2}\left(\dfrac{g_{0}}{g_{*}}\right)^{\frac{1}{3}}\\
		&\times\int_{0}^{\infty} dv \int_{|1-v|}^{1+v} du \left(\frac{4 v^{2}-\left(1+v^{2}-u^{2}\right)^{2}}{4 u v}\right)^{2}\\
		& \times
		P_{\mathcal{R}} (2\pi f u) P_{\mathcal{R}} (2\pi f v) I^{2}(u,v)\,,
	\end{split}
	\label{eq:final}
\end{equation}
where
\begin{equation}
	\begin{split}
		I^{2}(u,v)
		&= \dfrac{1}{2}\left(\frac{3}{4 u^{3} v^{3} x}\right)^{2}\left(u^{2}+v^{2}-3\right)^{2} \\
		&\times\Bigg\{ \left[-4 u v+\left(u^{2}+v^{2}-3\right) \ln \left|\frac{3-(u+v)^{2}}{3-(u-v)^{2}}\right|\right]^{2} \\
		&  +\left[\pi\left(u^{2}+v^{2}-3\right) \Theta(u+v-\sqrt{3})\right]^{2} \Bigg\}\,.
	\end{split}
	\label{eq:I3}
\end{equation}
$P_\mathcal{R}(k)$ is the power spectrum of curvature perturbations. We consider the case that $P_{\mathcal{R}}(k)$ has a power-law form around 
$k_{*}=2\pi f_*=20.6~\mathrm{pc}^{-1}$ (where $f_*=1~{\rm yr}^{-1}$),
\begin{equation}
	\label{eq:PR}
	P_{\mathcal{R}}(k)=P_{\mathcal{R}0}\left(\dfrac{k}{k_{*}}\right)^{m}\Theta(k-k_{\rm min})\Theta(k_{\rm max}-k)\,,
\end{equation}
$\Theta(x)$ is the Heaviside function. Note: to prevent overproduction of PBHs, we set cutoffs at $k_{\rm min}=0.03~k_{*}$ and $k_{\rm max}=100~k_{*}$ so that there is an upper bound on $P_{\mathcal{R}}(k)$. 

\noindent \textit{Prior parameters}:
\begin{itemize}
    \item $P_{\mathcal{R}0}$, the power spectrum of scalar perturbations at $k=k_{*}$;
    \item $m$, the slope of the power spectrum of scalar perturbations.
\end{itemize}

\noindent \textit{Other parameters}:
\begin{itemize}
    \item $g_{0}$, the present-day effective number of degrees of freedom of relativistic species;
    \item  $\Omega_{\mathrm{rad}}h^{2}=4.2\times 10^{-5}$, the present-day abundance of radiation.
\end{itemize}

\subsubsection{Plots of GW spectra}

The GW spectra from different sources confronting with the low-frequency five bins data of NANOGrav 15-yr, EPTA release-2 and PPTA DR3 are shown in Fig.~\ref{fig:median_fit}.

\section{Bayesian method}

Following Ref.~\cite{Bian:2020urb}, we briefly summarize here the Bayesian method for model comparisons and model fitting. 

Basically, Bayes' theorem gives
\begin{equation}
    P(M|D)= P(D|M) \frac{P(M)}{P(D)}
\end{equation}
where $D$ and $M$ respectively represent data and model. 
We also have
\begin{equation}
    P(D|M)=\int P(D|\theta,M) P(\theta|M) d\theta.
\end{equation}
To compare two models, $M_i$ and $M_j$, we calculate the ratio
\begin{equation}
    O_{ij} \equiv \frac{P(M_{i}|D)}{P(M_{j}|D)} = \frac{P(D|M_{i})}{P(D|M_{j})} \frac{P(M_{i})}{P(M_{j})} = B_{ij} \frac{P(M_{i})}{P(M_{j})}. 
\end{equation}
The ratio between the prior odds, $P(M_{i})/P(M_{j})$, is usually assumed to be close to 1. $B_{ij} \equiv P(D|M_{i})/P(D|M_{j})$ is the Bayes factor for $M_{i}$ vs. $M_{j}$.

The task of model fitting is to find the posterior distributions $P(\theta|D,M)$ of parameters $\theta$ for known data $D$ and model $M$. Again, Bayes' theorem gives
\begin{equation}
    P(\theta|D,M) = \frac{P(D|\theta,M) P(\theta|M)}{P(D|M)},
\end{equation}
The conditional probabilities $P(D|\theta,M)$,  $P(\theta|M)$, and $P(D|M)$ are, respectively, the likelihood, prior, and, model evidence.

As discussed in the main text, we have presented the model comparisons in eqs.~(\ref{eqB})-(\ref{eqB_epta}) based on Bayes factors. We have utilized the datasets of NANOGrav 15-yr dataset, EPTA release-2, and PPTA DR3 and the SGWB models of SMBHB, FOPT, cosmic strings, domain walls, and large amplitude curvature perturbations.
The results of model fitting have also been shown in Figs.~\ref{SMBHB}-\ref{IGW}, with the help of \texttt{emcee} and \texttt{GetDist}~\cite{Foreman-Mackey:2012any,Lewis:2019xzd}.

\begin{table*}[htbp]
\caption{Model parameters and their prior distribution in data analysis. U and log-U stand for the uniform and log-uniform distribution.}
\label{PPT}
\centering\setlength{\tabcolsep}{8mm}
\begin{tabular}{lcc}
\hline\hline
\textbf{Parameter}         & \textbf{Description}     & \textbf{Prior}  \\ 
\hline
~&\textbf{SMBHB}&~\\
$A$          & amplitude & log-U$[-18, -12]$                                \\
\hline
~&\textbf{FOPT}&~\\
$\beta/H_{*}$          & inverse PT duration & U$[5,70]$        \\
$T_{*}$          & PT temperature & U$[0.01,1.6]$          \\
$\alpha_{PT}$          & PT strength & U$[0.0,1.0]$            \\
$v_{b}$          & velocity of bubble wall & U$[0.9,1.0]$      \\
\hline
~&\textbf{Cosmic string}&~\\
$G\mu$          & string tension & log-U$[-12, -6]$ \\
$\alpha_{CS}$          & loop-size parameter & log-U$[-6, 0]$ \\
\hline
~&\textbf{Domain wall}&~\\
$\sigma$          & surface energy density & log-U$[0,8]$ \\
$\Delta V$          & bias potential & log-U$[0,8]$ \\
\hline
~&\textbf{Curvature perturbations}&~\\
$P_{\mathcal{R}0}$          & amplitude of spectrum & log-U$[-4, -1.5]$ \\
$m$          & slope of spectrum  & U$[-2, 2]$                                \\
\hline\hline
\end{tabular}
\end{table*}

\end{document}